\documentclass{iaus}
\usepackage{graphicx}

\title[RMHD protostellar collapse calculations] %% give here short title %%
{Radiative, magnetic and numerical feedbacks on small-scale fragmentation}

\author[Beno\^it Commer\c con et al.]   %% give here short author list %%
{Beno\^it Commer\c con$^{1}$, Patrick Hennebelle$^2$, Edouard Audit$^3$, Gilles Chabrier$^4$  \and Romain Teyssier$^{3,5}$}

\affiliation{$^1$Max PLanck Institut f\"ur Astronomie \\ 
K\"onigstuhl 17, D-69117 Heidelberg, Germany \\ email: {\tt benoit@mpia-hd.mpg.de} \\[\affilskip]
$^2$Laboratoire de radioastronomie, \'Ecole Normale Sup\'erieure et Observatoire de Paris, \\
24 rue Lhomond, F-75231 Paris Cedex 05, France \\[\affilskip]
$^3$Laboratoire AIM, CEA/DSM - CNRS - Universit\'e Paris Diderot,\\ 
IRFU/SAp, F-91191 Gif sur Yvette, France \\[\affilskip]
$^4$\'Ecole Normale Sup\'erieure de Lyon, Centre de recherche Astrophysique de Lyon, \\
46 all\'ee d'Italie, F-69364 Lyon Cedex 07, France \\[\affilskip]
$^5$Universit\"at Z\"urich, Institute f\"ur Theoretische Physik, \\
Winterthurerstrasse 190, CH-8057 Z\"urich, Switzerland}

\pubyear{2010}
\volume{270}  %% insert here IAU Symposium No.
\pagerange{0-0}
\jname{Computational Star Formation}
\editors{Alves, Elmegreen, Girart, Trimble, eds.}

\begin{document}

\maketitle

\begin{abstract}
{Radiative feedback and magnetic field are understood to have a strong impact on the protostellar collapse. We present high resolution numerical calculations of the collapse of a 1 M$_\odot$ dense core in solid body rotation, including both radiative transfer and magnetic field. Using typical parameters for low-mass
  cores, we study thoroughly the effect of radiative transfer and magnetic field on the first core formation and fragmentation. We show that including the two aforementioned physical processes does not correspond to the simple picture of adding them separately. The interplay between the two is extremely strong, via the magnetic braking and the radiation from the accretion shock.}
\end{abstract}

\firstsection % if your document starts with a section,
\section{Introduction}

The protostellar collapse of low mass dense cores follows a well-defined sequence of different stages, down to the formation of a protostar. In the first collapse phase \cite[(Larson 1969)]{Larson_69}, the compressed gas cools efficiently thanks to the coupling between gas and dust. At higher densities ($\rho>10^{-13}$ g cm$^{-3}$), the radiation is trapped and the first hydrostatic core (the first Larson core) is formed. At this stage, the grain opacities and the radiation transport play a major role. In the recent past few years, a lot of progress in the computational star formation field has be done. For instance, a lot of radiation hydrodynamics (RHD) methods have been developed for grid based codes (e.g., \cite[Krumholz et al. 2007]{Krumholz_07}, \cite[Kuiper et al. 2010]{Kuiper_2010}) and for smoothed particles hydrodynamics (SPH) codes (\cite[Whitehouse \& Bate 2006]{Whitehouse_Bate_06}, \cite[Stamatellos et al. 2007]{Stamatellos_07}).
Applying these methods to star formation, it turns out that a barotropic EOS cannot account for realistic cooling and heating of the gas \cite[(e.g., Commer\c con et al. 2010]{Commercon_10}). On larger scales, radiative transfer has been found to efficiently reduce the fragmentation thanks to radiative feedback due to the accretion and the protostellar evolution (\cite[Bate 2009]{Bate_09}, \cite[Offner et al. 2009]{Offner_09}).
Regarding magnetic field in the star formation context, a gradually improved expertise has been developed for magnetohydrodynamical (MHD) flows (e.g., \cite{Hennebelle_08}, \cite{Machida_08}). All these studies showed that magnetic fields reduce efficiently the fragmentation of prestellar cores. 
Recently, it has been shown that both radiative transfer and ideal MHD are important and cannot be neglected in the collapse and fragmentation of
  protostellar cores (\cite[Price \& Bate 2009]{Price_09}, \cite[Commer\c con et al. 2010]{Commercon_10}, \cite{Tomida_10}).

In this study, we present  radiation magnetohydrodynamics (RMHD) calculations of prestellar dense core collapse, using the adaptive mesh refinement (AMR) code {\ttfamily RAMSES} \cite[(Teyssier 2002)]{Teyssier_02}. 
The paper is organized as follows:  in section 2, we briefly introduce the RMHD solver we designed and present our initial conditions. In section 3, we present our results of dense core collapse calculations using the RMHD solver with various numerical resolutions. Finally, we draw our conclusion in section 4.

\section{Numerical method and initial conditions}

We use the {\ttfamily RAMSES} code \cite[(Teyssier 2002)]{Teyssier_02}, in which we have implemented a RHD solver using the grey flux-limited diffusion \cite[(FLD, e.g. Minerbo 1978)]{Minerbo_78} approximation. We use the comoving frame, which is valid to leading order in $v/c$ in the static diffusion and streaming limits.
The solver we designed is coupled to the ideal magneto hydrodynamics one developed by \cite[Fromang et al. (2006)]{Fromang_06}. The coupled RMHD equations read
 \begin{equation}
\left\{
\begin{array}{llllll}
\partial_t \rho & + & \nabla \left[\rho\textbf{u} \right] & = & 0 \\
\partial_t \rho \textbf{u} & + & \nabla \left[\rho \textbf{u}\otimes \textbf{u} - \textbf{B}\otimes\textbf{B}+ P \mathbb{I} \right]& =& -\rho\nabla\Phi - \lambda\nabla E_\mathrm{r} \\
\partial_t E_\mathrm{T} & + & \nabla \left[\textbf{u}\left( E_\mathrm{T} + P_\mathrm{tot}\right) - \textbf{B}(\textbf{B}.\textbf{u})  \right] &= &-\rho\textbf{u}\cdot\nabla \Phi - \mathbb{P}_\mathrm{r}\nabla:\textbf{u}  - \lambda \textbf{u} \nabla E_\mathrm{r} \\
 & & & &+  \nabla \cdot\left(\frac{c\lambda}{\rho \kappa_\mathrm{R}} \nabla E_\mathrm{r}\right) \\
\partial_t E_\mathrm{r} & + & \nabla \left[\textbf{u}E_\mathrm{r}\right]
&=& 
- \mathbb{P}_\mathrm{r}\nabla:\textbf{u}  +  \nabla \cdot\left(\frac{c\lambda}{\rho \kappa_\mathrm{R}} \nabla E_\mathrm{r}\right) \\
 & & & & + \kappa_\mathrm{P}\rho c(a_\mathrm{R}T^4 - E_\mathrm{r})\\
\partial_t \textbf{B}  & - &  \nabla \times (\textbf{u}\times\textbf{B})&=&0, 
\end{array}
\right.
\end{equation}
where $\rho$ is the density, $\textbf{u}$ is the velocity vector, $\textbf{B}$ is the magnetic field vector,  $P_\mathrm{tot}$ is  the total pressure, sum of the thermal and magnetic pressures, $P_\mathrm{tot}=P +  \frac{\textbf{B}.\textbf{B}}{2}$, $E_\mathrm{r}$ is the radiative energy, $\lambda$ is the flux limiter \cite[(Minerbo 1978)]{Minerbo_78}, 
$E_\mathrm{T}$ is the total fluid energy per unit volume,
$E_\mathrm{T}=\rho(\epsilon + \frac{\textbf{u}.\textbf{u}}{2})+ E_\mathrm{r}+\frac{\textbf{B}.\textbf{B}}{2}$,
$\Phi$ is the gravitational potential, $\mathbb{P}_\mathrm{r}$ is the radiative pressure, and $\kappa_\mathrm{R}$ and $\kappa_\mathrm{P}$ are the Rosseland and Planck mean opacities. 
This system of equations is closed by the perfect gas equation of state $P/\rho=(\gamma-1)\epsilon$, where $\gamma=5/3$ is the adiabatic exponent. An additional constraint comes from the divergence of the magnetic field, which has to vanish everywhere at all times ($\nabla. \textbf{B} =0$). Note that the radiation diffusion and coupling terms are integrated implicitly in time, since it involves very short timescale processes, compared to the hydrodynamical evolution.  

\textit{- Initial conditions:} 
We     consider    a uniform-density  sphere  of molecular  gas, rotating  about   the  $z$-axis   with  a  uniform   angular  velocity. The prestellar core mass is fixed at $M_0 =  1$ M$_{\odot}$  and the temperature at 11 K, which corresponds to an isothermal sound speed $c_{\mathrm{s}0}   \sim  0.19   $  km s$^{-1}$.    To promote fragmentation,  we  add  an $\mathrm{m}=2$  azimuthal  density perturbation with an amplitude of 10\%.
The magnetic field is initially uniform and parallel to the rotation axis. The strength of the magnetic field is expressed in terms of the mass-to-flux to critical mass-to-flux ratio $\mu=20=(M_0/\Phi)/(M_0/\Phi)_\mathrm{c}$.
The  initial  ratio of  the
thermal to gravitational energies is $\alpha =0.37$,  and the initial ratio of the rotational to gravitational energies is $\beta=0.045$.

Calculations were performed using
either the  rather diffusive Lax Friedrich (LF) Riemann solver or the more accurate HLLD Riemann solver (\cite{Miyoshi_Kusano_05}). Following up on former studies (\cite{Commercon_08}), we impose at least 15 cells per Jeans length as a grid refinement criterion (parameter $N_\mathrm{J}$). We also use one additional parameter, $N_\mathrm{exp}$, which indicates the number of cells refined in each direction around a cell violating the Jeans length criterion. 
 The initial resolution of the grid contains $64^3$ cells.
We  use the low temperature grey opacities of \cite{Semenov_03}.

\begin{figure}[t]
% \vspace*{-2.0 cm}
\begin{center}
 \includegraphics[width=8.6cm]{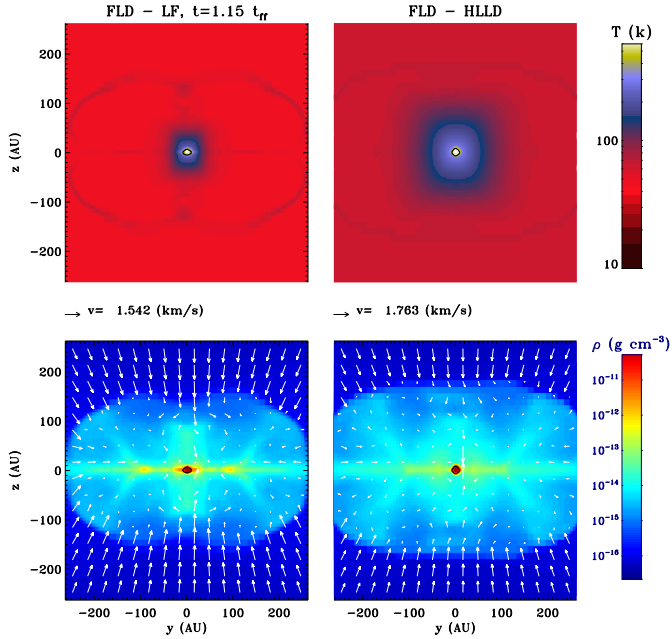} 
% \vspace*{-1.0 cm}
 \caption{Temperature and density maps in the {\it yz}-plane at time $t=1.15 t_\mathrm{ff}$ for two calculations with the FLD and the LF (left) and the HLLD (right) Riemann solvers.}
   \label{fig1}
\end{center}
\end{figure}

\section{Results}

{- \it Effect of the solver:} Figure \ref{fig1} shows temperature and density maps in the {\it yz}-plane for two calculations with the FLD and the LF  and the HLLD Riemann solvers. 
 Each calculation has been performed using $N_\mathrm{J}=15$ and $N_\mathrm{exp}=4$. 
The FLD-LF case leads to spurious fragmentation as it is shown in \cite{Commercon_10}. 
The bubble (dense region, $\rho>10^{-15}$ g cm$^{-3}$), driven by magnetic pressure due to the magnetic field line wrapping, is less extended in the FLD-LF case, and the disc is thus more massive and more prone to fragmentation. We identify the accretion shock on the first Larson core as a {\it supercritical radiative shock}, i.e. all the infalling kinetic energy is radiated away.
Consequently, the radiative feedback due to the accretion on the first Larson core is much larger in the FLD-HLLD case, since the magnetic braking and thus the infall velocity are larger thanks to the less diffusive HLLD Riemann solver \cite[(Commer\c con et al. 2010)]{Commercon_10}.

\begin{figure}[t]
% \vspace*{-2.0 cm}
\begin{center}
 \includegraphics[width=13cm,height=8.25cm]{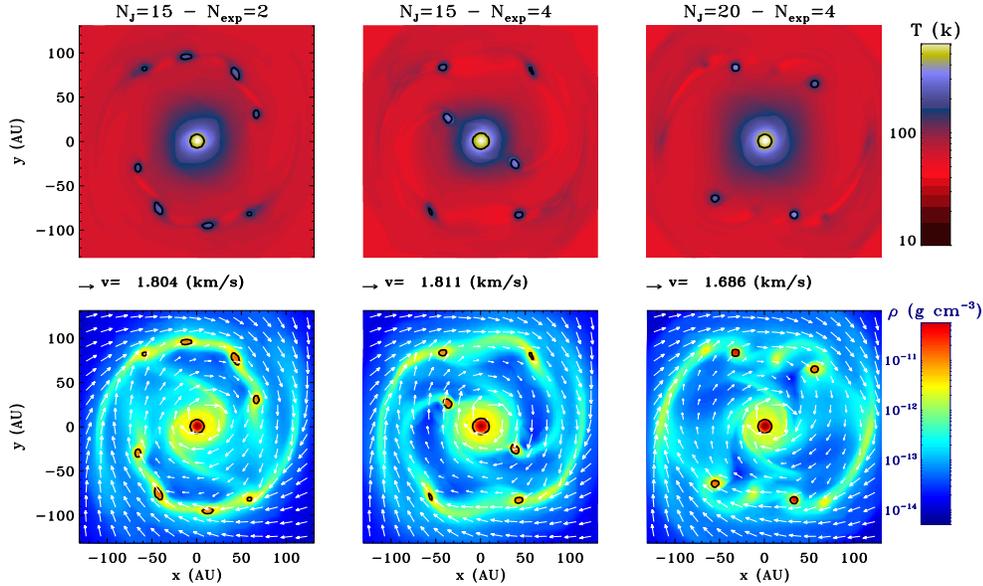} 
% \vspace*{-1.0 cm}
 \caption{Temperature and density maps in the {\it xy}-plane  at time $t=1.15 t_\mathrm{ff}$ for three calculations using the FLD and the LF Riemann solver, with various resolutions.}
   \label{fig2}
\end{center}
\end{figure}

{- \it Effect of the numerical resolution: } Figure \ref{fig2} shows temperature and density maps in the {\it xy}-plane for three calculations using the FLD and the LF Riemann solver, with various resolutions: 
($N_\mathrm{J}=15$; $N_\mathrm{exp}=2$), ($N_\mathrm{J}=15$; $N_\mathrm{exp}=4$) and ($N_\mathrm{J}=20$; $N_\mathrm{exp}=4$). We clearly see that increasing the resolution, from left to right, leads to a decrease in the number of fragments produced. As resolution increases, the diffusivity of the LF solver is reduced, the disk is then less massive, and the magnetic braking more efficient.  Using the HLLD solver and barotropic calculations, \cite{Commercon_10} show that the correct behavior is the case without fragmentation.

\section{Conclusion}

We show that taking into account both radiative transfer and magnetic field is not a straightforward linear process. We show that the magnetic braking and the magnetic bubble extent influence: i) the radiative feedback via the infall velocity and ii) the fragmentation via the disc mass and the rotational velocity. 
Last but not least, the results are extremely sensitive to the numerical resolution and to the numerical diffusivity of the code used, which readers and authors should be aware of.

\end{document}